\documentstyle[prl,twocolumn,aps,epsfig]{revtex}
\begin{document}

\draft
\wideabs{
\title{Observation of Phase Fluctuations in Bose-Einstein Condensates}

\author{S. Dettmer, D. Hellweg, P. Ryytty, J. J. Arlt, W. Ertmer,
and K. Sengstock}
\address{Institut f\"ur Quantenoptik, Universit\"at Hannover, Welfengarten 1,
          30167 Hannover, Germany}
\author{D. S. Petrov$^{1,2}$ and G. V. Shlyapnikov$^{1,2,3}$}
\address{$^1$FOM Institute for Atomic and Molecular Physics,
        Kruislaan 407, 1098 SJ Amsterdam, The Netherlands \\
        $^2$Russian Research Center Kurchatov Institute, Kurchatov Square, 123182 Moscow, Russia \\
        $^3$Laboratoire Kastler Brossel, Ecole Normale Sup$\acute e$rieure, 24 rue Lhomond, 75231 Paris Cedex 05, France}
\author{H. Kreutzmann, L. Santos and M. Lewenstein}
\address{Institut f\"ur Theoretische Physik,
Universit\"at Hannover, Appelstra\ss e 2,
          30167 Hannover, Germany}
\date{\today}
\maketitle

\begin{abstract}
\noindent

The occurrence of phase fluctuations due to thermal excitations in Bose-Einstein
condensates (BECs) is studied for a variety of temperatures and trap geometries.
We observe the statistical nature of the appearence of phase fluctuations and
characterize the dependence of their average value on
temperature, number of particles and the trapping potential.
We find pronounced phase fluctuations for condensates in very elongated traps in a broad
temperature range. The results are of great importance for the realization of BEC in
quasi 1D geometries, for matter wave interferometry with BECs, as well as
for coherence properties of guided atom laser beams.
\end{abstract}

}

Since the first experimental realization of Bose-Einstein condensation
in dilute atomic gases \cite{BECs}, there has been enormous interest
in the coherence properties of BECs. In particular, the phase coherence is
essential for applications of BEC in matter wave interferometry;
it also sets limits on the coherence of atom lasers,
and guided atom laser beams.
For a trapped 3D condensate well below the BEC transition temperature $T_{c}$,
recent experiments have
confirmed the phase coherence, e.g., it was shown that the coherence length is
equal to the condensate size \cite{coherenceMIT,coherenceNIST}.
However, phase coherence is not an obvious property of BEC. In particular,
it is expected that low-dimensional (1D and 2D) quantum gases
differ qualitatively from the 3D case in this respect
\cite{Shlyapnikov1D,Shlyapnikov2D,Kagan2D,exp_quasi2D}.
Recently, it was shown theoretically \cite{Shlyapnikov3D} that
for very elongated condensates
phase fluctuations can be pronounced already in the equilibrium state
of the usual 3D ensemble, where the density fluctuations are suppressed.
The phase coherence length in this case can be smaller than
the axial size of the sample. This is referred to as the regime
of quasicondensation \cite{remark}.
The detailed characterization of
phase fluctuations in condensates is thus of great importance for
applications of BEC, especially for recent attempts
to reach BEC in elongated micro-circuit geometries \cite{micro}.
The temperature dependence of the coherence of an atom laser beam
was studied in \cite{coherenceM}.

In this Letter we report on systematic studies of BEC of
$^{87}$Rb atoms in the regime of phase-fluctuating condensates.
We achieved this regime
in highly  anisotropic traps leading to a strongly elongated shape
of the condensate.
We observe the phase fluctuations by measuring the density distribution of
the released cloud after ballistic expansion.
 By varying the temperature and the aspect ratio of the trapping potential,
 we study the continuous transition from the usual 3D regime, where phase
 fluctuations of the condensate are low, into the regime of strong phase
 fluctuations.
We show that a phase coherent matter wave is not a direct outcome of BEC
in strongly elongated geometries but can only be achieved for very low
temperatures, well below the BEC transition temperature $T_{c}$.

Fluctuations of the phase of a Bose condensate are related to thermal excitations
and always appear at finite temperature. However, as shown in
Ref.~\cite{Shlyapnikov3D}, the fluctuations depend not only
on temperature but also on the trap geometry, and on the particle number.
Typically, fluctuations
in spherical traps are strongly suppressed as the wavelengths of excitations are
smaller than the size of the atomic cloud.
In contrast, wavelengths of the excitations in strongly elongated traps can be larger
than the {\it radial} size of the cloud. In this case, the low-energy {\it axial} excitations
acquire a 1D character and can lead to more pronounced phase fluctuations,
although the density fluctuations of the equilibrium state are still
suppressed.
Due to the pronounced phase fluctuations the coherence properties of elongated
condensates can be significantly altered as compared with previous observations.
In particular, the axial coherence length can be much smaller than the
size of the condensate, which can have dramatic consequences for practical applications.

The experiment was performed with Bose-Einstein condensates of up to
$N_{0}\!=5\!\times\! 10^{5}$ rubidium atoms in the $|F\!=\!2,
m_{F}\!=\!+2\rangle $ state of a cloverleaf-type magnetic trap. Further details
of our apparatus were described previously \cite{bouncing,Solitons}. The
fundamental frequencies of the magnetic trap are $\omega_{x}=2\pi\times 14\,
$Hz and $\omega_{\rho}=2\pi\times 365\, $Hz along the axial and radial
direction, respectively. Due to the highly anisotropic confining potential with
an aspect ratio $\lambda\!=\omega_{\rho}/\omega_{x}$ of 26, the condensate is
already elongated along the horizontal $x$ axis. In addition, further radial
compression of the ensemble by means of a superimposed blue detuned optical
dipole trap is possible \cite{Waveguide}. We performed the measurements for
radial trap frequencies $\omega_{\rho}$ between $2\pi\times 138\, $Hz and
$2\pi\times 715\, $Hz corresponding to aspect ratios $\lambda$ between 10 and
51. After rf evaporative cooling to the desired temperature, we wait for $1\,
$sec (with rf 'shielding') to allow the system to reach an equilibrium state.
We then switch off the trapping potential within $200\, \mu $s and allow a
variable time-of-flight.

\begin{figure}[h]
   \begin{center}
   \parbox{8cm}{\epsfxsize 8cm\epsfbox{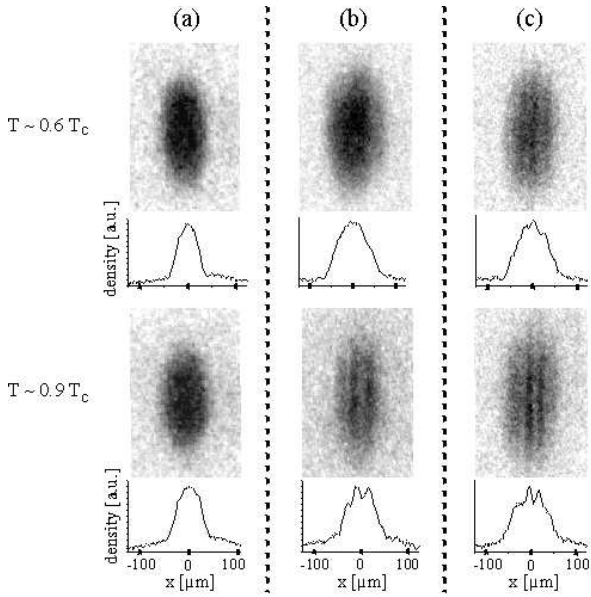} }
   \end{center}
   \caption{Absorption images and corresponding density profiles of
            BECs after $25\, $ms time-of-flight taken for
            various temperatures $T$ and aspect ratios
            [$\lambda = 10$ (a), 26 (b), 51 (c)].}
   \label{fig:TOF}
\end{figure}

Figure \ref{fig:TOF} shows examples of experimental
data for various temperatures $T<T_{c}$ and aspect ratios $\lambda$.
The usual anisotropic expansion of the condensate  related to the
anisotropy of the confining potential is clearly visible in the absorption images.
The line density profiles below reflect the parabolic shape of
the BEC density distribution.
Remarkably, we observe pronounced stripes in the density distribution in some cases
shown in Fig.~\ref{fig:TOF}.
On average these stripes are more pronounced at high aspect ratios of the trapping potential
[Fig.~\ref{fig:TOF}(c)], high temperatures (bottom row of Fig.~\ref{fig:TOF}), and
low atom numbers.

The appearence of stripes can be understood qualitatively as follows. As mentioned above,
within the equilibrium state of a BEC in a magnetic trap the density distribution remains
largely unaffected even if the phase fluctuates \cite{Shlyapnikov3D}.
The reason is that the mean-field interparticle interaction prevents the transformation
of local velocity fields provided by the phase fluctuations into modulations of the
density.
However, after switching off the trap, the mean-field interaction rapidly decreases
and the axial velocity fields are then converted into the density distribution.
We have performed numerical simulations of the 3D Gross-Pitaevskii equation
to understand quantitatively how phase fluctuations lead to the build up of stripes
in the density distribution. We assume that initially the condensate had an
equilibrium density profile, and a random fluctuating phase $\phi(x)$.
For elongated BECs \cite{Shlyapnikov3D} the phase can be  represented as
$\phi(x)=\sum_{j=1}^{\infty}\phi_j(x)$, where
\begin{equation}\label{phi_j}\!\phi_j(x)\!=\!\left [ \frac{(j\!+\!2)(2j\!+\!3)g}
{4\pi R^2 L \epsilon_j (j\!+\!1)} \right ] ^{1/2}\!\!\!
P_j^{(1,1)}\left(\frac{x}{L}\right)
 \frac{(\alpha_j+\alpha_j^{\ast})}{2}.
\end{equation}
Here  $\epsilon_j=\hbar\omega_x\sqrt{j(j+3)/4}$ is the spectrum of low-energy
axial excitations \cite{Stringari}, $P_j^{(1,1)}$ are Jacobi polynomials, $g=4\pi\hbar^2
a/m$, $a>0$ is the scattering length, and $R$ ($L$) is the radial (axial)
condensate size. In Eq.~(\ref{phi_j}),  quasiparticle creation and
annihilation operators have been replaced by complex amplitudes $\alpha_j $ and
$\alpha_j^{\ast}$. To reproduce the quantum statistical  properties of the
phase,  $\alpha_j $ and $\alpha_j^{\ast}$ were sampled as random variables
with a zero mean value and $\langle |\alpha_j|^2 \rangle =N_j$,
where $N_j=[{\rm exp}(\epsilon_j/k_{B}T)-1]^{-1}$ is
the occupation number for the quasiparticle mode $j$.

\begin{figure}[h]
   \begin{center}
   \parbox{6.5cm}{\epsfxsize 6.5cm\epsfbox{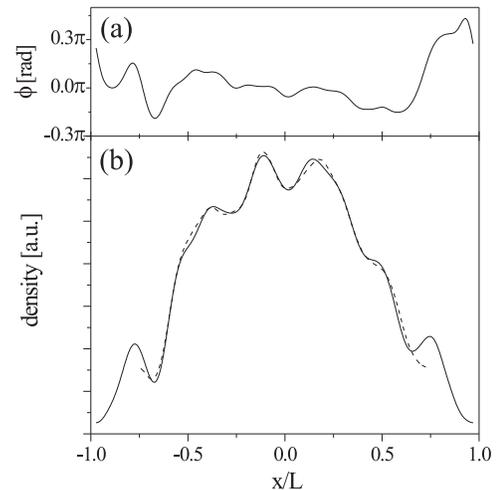} }
   \end{center}
   \caption{(a) A typical initial phase distribution calculated
                for $\omega_x=2\pi\times 14\, $Hz, $\omega_{\rho}=2\pi\times 508\, $Hz,
                $N_0=2\times 10^5$, and $T=0.5\, T_{c}$.
            (b) Corresponding density profile after $25\, $ms time-of-flight
                from simulations (solid line) and analytical theory (dashed line).}
   \label{fig:figth}
\end{figure}

Alternatively, the formation of the stripes has been studied by using the local
density approximation for the axial profile and by relying on the scaling
approach \cite{selfsimilar} for the radial expansion of the cloud.
The rescaled Gross-Pitaevskii equation was then linearized with respect to
small fluctuations of the phase gradients and the density. Our method accounts for
the transformation of initially phonon-like excitations into particle-like
ones and covers both the (initially) hydrodynamic and the (ultimately)
free-particle regimes of expansion.
For a condensate which initially is in the Thomas-Fermi regime and has a chemical potential $\mu $,
the radially integrated density fluctuations $\delta n(x)$ after a time-of-flight $t$ in the
interval $\mu/\hbar\omega_x^2\gg t\gg\mu/\hbar\omega_{\rho}^2$ are given by
\begin{equation}  \label{delta}
\frac{\delta n(x)}{n_{0}(x)}\!=\! 2\!\sum_j \sin\!
\left[\frac{\epsilon_j^2t}{\hbar\mu
(1\!-\!(\frac{x}{L})^2)}\right]\!(\omega_{\rho}t)^{-(\epsilon_j/\hbar\omega_\rho)^2}\!\phi_j(x).
\end{equation}
Here the profile $n_{0}(x)$ is the radially integrated density for the unperturbed
condensate.
From Eq.~(\ref{delta}) one obtains a closed relation for the mean square density
fluctuations $(\sigma_{\mbox{\tiny BEC}}/n_{0})^2$ by averaging $(\delta n/n_0)^2$ over different
realizations of the initial phase. In the central part of the cloud ($x\approx 0$)
we find
\begin{equation}
\!\!\left ( \!\frac{\sigma_{\mbox{\tiny BEC}}}{n_{0}}\! \right )^2\!=\!\frac{T}{\lambda
T_{\phi}}\sqrt{\frac{\ln\tau}{\pi}}
\left (
\sqrt{\!
1\!+\!
\sqrt{\!
1\!+\!
\left ( \frac{\hbar\omega_\rho\tau}{\mu\ln\tau} \right )^2
}
}
\!-\!\sqrt{2}
\right )\!,\!\!
\label{therm}
\end{equation}
where $\tau=\omega_\rho t$. In Eq.~(\ref{therm}) we introduced a characteristic
temperature $k_{B}T_\phi=15(\hbar\omega_x)^2N_0/32\mu$.
For $T_{\phi}<T_c$ one expects the regime of quasicondensation
for the initial cloud in the temperature interval $T_{\phi}<T<T_c$,
i.e., the regime where the coherence length is smaller than the condensate size \cite{Shlyapnikov3D}.
The analytical expressions agree very well with the numerical
simulations  (see Fig.~\ref{fig:figth}).  Note that Eq.~(\ref{therm})
provides a direct relation between the observed density fluctuations and the
temperature, and thus can be used for thermometry at very low $T$.

\begin{figure}[h]
   \begin{center}
   \parbox{7.5cm}{\epsfxsize 7.5cm\epsfbox{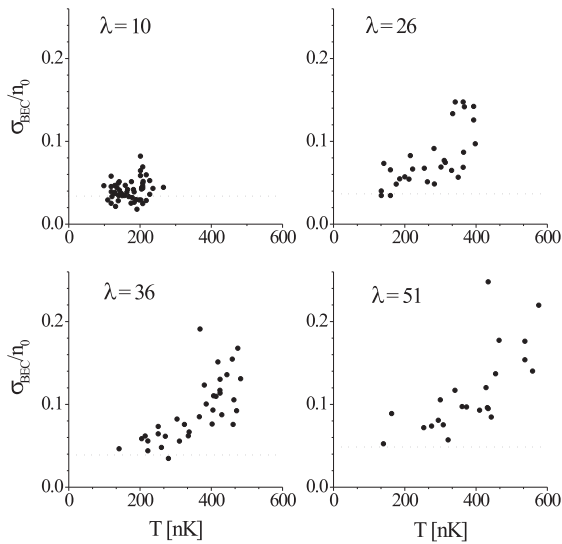} }
   \end{center}
   \caption{Measurement of $\sigma_{\mbox{\tiny BEC}}/n_{0}$ versus temperature
            in four different trap geometries.
            The dotted lines represent the average detection noise $\sigma_{T}/n_{0}$.
            All data was taken for $T<T_{c}$.}
   \label{fig:AR}
\end{figure}

To determine the amount of phase fluctuations experimentally, we systematically studied the
formation and structure of stripes in the atomic density distribution as a function of the
trapping potential and temperature.
For each realization of a BEC, the observed density distribution was integrated along the
radial direction and then fitted by a bimodal function with the integrated parabolic Thomas-Fermi
distribution for the condensate fraction and a Gaussian for the thermal cloud.
For each image we obtained standard deviations $\sigma_{\mbox{\tiny BEC}}$ of the experimental
data from the fit in the central region of the condensate fraction (half width of full size).

The temperature dependence of $\sigma_{\mbox{\tiny BEC}}$ is shown in Fig.~\ref{fig:AR}
for four different trap configurations with aspect ratios of $\lambda=10, 26, 36$ and $51$.
To account for shot-to-shot variations in the atom number, the standard deviations
were normalized to the fitted peak density $n_{0}$ in the condensate.
The dotted lines show the average detection noise $\sigma_T/n_{0}$
for data from each trap, obtained from the thermal wings of the cloud.

Since the initial phase of a Bose condensate is mapped into
its density distribution after time-of-flight, the quantity
$\sigma_{\mbox{\tiny BEC}}$ is a direct measure of the initial phase fluctuations.
Note, however, that this method reflects the instantaneous phase of the BEC at the
time of release and, therefore, images taken at the same initial conditions can look
significally  different.  Indeed, we observe a large spread of our
experimental data (see Fig.~\ref{fig:AR}),
which clearly demonstrates the statistical character of the phase fluctuations.
Furthermore, our data characterizes phase fluctuations for various trap geometries.
For $\lambda =10$ the data falls into our detection noise and, therefore, the phase
fluctuations are hardly observable.
Considerably
different images are observed for more elongated traps  with
$\lambda =26, 36$ and $51$.
In these cases we observe significant deviations
from the Thomas-Fermi distribution at high temperatures, which is a clear
indication of the presence of phase fluctuations.
For lower temperatures, reached by further evaporation, phase fluctuations
become reduced due to both, the reduced excitation spectrum at lower temperature
and the increasing number of atoms in the condensate fraction.
However, for even further evaporation the number of Bose-condensed atoms $N_0$
decreases leading to phase fluctuations in agreement with Eq.~(\ref{therm}).
In fact, it was very difficult to reduce the density modulation below the noise limit
by rf evaporation in the case of our tightest trap with $\lambda=51$.

As observed in our experiment phase fluctuations for BECs in elongated
traps continuously decrease with reducing $T$ and increasing $N_0$.
It is not possible to determine a cut-off for the phase
fluctuations, rather they  decrease until they cannot be resolved below our
noise limit. Hence all experiments with BECs at finite temperature in tightly
confining elongated potentials will be subject to axial phase fluctuations.

\begin{figure}[h]
   \begin{center}
   \parbox{7.5cm}{\epsfxsize 7.5cm\epsfbox{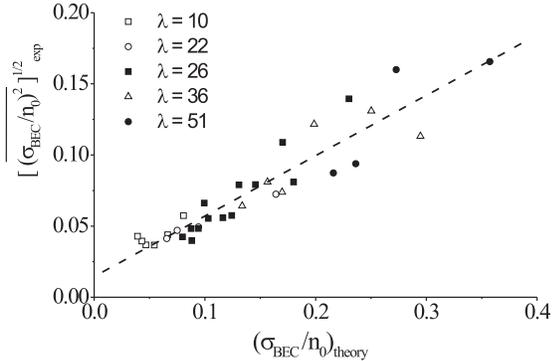} }
   \end{center}
   \caption{Average standard deviation of the measured line densities
            $[\, \overline{(\sigma_{\mbox{\tiny BEC}}/n_{0} )^2}\, ]^{1/2}_{\mbox{\tiny exp}}$
            compared to the theoretical value of
            $(\sigma_{\mbox{\tiny BEC}}/n_0)_{\mbox{\tiny theory}}$
            obtained from  Eq.~(\ref{therm}).
            The dashed line is a fit to the experimental data.}
   \label{fig:sigma}
\end{figure}

In order to obtain general information about the phase fluctuations, we average the
observed standard deviations over many realizations within a small temperature and particle number interval.
This allows us to use Eq.~(\ref{therm}) and to compare directly the averaged measured value of
$[\, \overline{(\sigma_{\mbox{\tiny BEC}}/n_{0} )^2}\, ]^{1/2}_{\mbox{\tiny exp}}$
to the predicted value of $(\sigma_{\mbox{\tiny BEC}}/n_0)_{\mbox{\tiny theory}}$
for various values of $T, N_0, \omega_x, \omega_{\rho}$, and $t$ (see Fig.~\ref{fig:sigma}).
The theoretical value takes also the limited experimental imaging resolution into account.

Our experimental results follow the expected general dependence very well.
With the direct link of the phase fluctuations in the magnetic trap to the observed
density modulation given by Eq.~(\ref{therm}), our data therfore confirms the predicted
general behavior of phase fluctuations in elongated BECs.
However, the measured values are approximately by a factor of 2 smaller than
those predicted by theory.
This discrepancy could be due to, e.g., a reduction of the observed contrast caused by a
small tilt in the detection laser beam with respect to the radial stripes.
Most of the observed experimental
data,  which exhibit fluctuations well above the noise limit, correspond to $T>T_{\phi}$.
This implies that the measurements
were performed in the regime of quasicondensation, i.e., the phase coherence
length $l_{\phi}=LT_{\phi}/T$ of the initial condensate was smaller than the axial size $L$.
For instance, for $\lambda=51$, $T=0.5\, T_{c}$, and $N_{0}=3\times 10^4$, one obtains
$l_{\phi}\approx L/3$.

In conclusion, we have presented detailed experimental and theoretical studies of
a BEC state with fluctuating phase. The flexible method of
ballistic expansion allows to measure phase fluctuations under various experimental
conditions, especially for various trap geometries.
Bragg spectroscopy can also be used to measure the influence of phase fluctuations
on the momentum distribution of particles in the axial direction, but this method is
difficult to apply to very elongated traps.
In addition, the ballistic expansion for very long times can visualize extremely small
phase fluctuations not accessible to the usual resolution of Bragg spectroscopy.
By measuring the phase fluctuations and comparing the temperature with $T_{\phi}$
we have demonstrated instances, where the phase coherence length was smaller than
the axial size of the condensate, i.e., the initial cloud was in the
quasicondensate state.  Our results set severe limitations on applications
of BECs in interferometric measurements, and for guided atom laser beams.
Our experimental method combined with the theoretical
analysis [Eq.~(\ref{therm})] provides a  method of BEC thermometry.
Further studies of quasicondensates, e.g., with respect to
superfluid properties, will be necessary to obtain a full understanding
of the phase coherence properties of ultracold atomic gases.

This work is supported by SFB\,407 of the {\it Deutsche Forschungsgemeinschaft}.
DSP and GVS acknowledge support from the Alexander von Humboldt Foundation,
from the Dutch Foundations NWO and FOM, and from the Russian Foundation for
Basic Research.


\begin{references}
\bibitem{remark} Here, we refer to the equilibrium state of the system, which has a fluctuating phase,
                 in contrast to the notion of a 'quasicondensate' appearing as an intermediate state
                 during the formation process of a BEC \cite{Kagan,Ketterle_quasi}.
\end{references}
\end{document}